\documentclass[onecolumn,showpacs,preprintnumbers,prl,superscriptaddress]{revtex4}
\usepackage{amsfonts,amssymb,amsmath,latexsym,epsfig} 
\usepackage{color}
\usepackage{hyperref}

\newcommand{\rz}{{\mathbb R}}

\begin{document}

\preprint{}
\title{Sparre-Andersen theorem with spatiotemporal correlations}
\author{Roberto Artuso}
 \email{roberto.artuso@uninsubria.it}
\affiliation{%
Center for Nonlinear and Complex Systems, Universit\`a dell' Insubria, Via Valleggio 11, Como, 22100, Italy
}
\affiliation{%
I.N.F.N Sezione di Milano, Via Celoria 16, Milano, 20133, Italy
}%

\author{Giampaolo Cristadoro}
 \email{giampaolo.cristadoro@unibo.it}
\affiliation{%
Dip. Matematica, Universit\`a di Bologna, Piazza di Porta San Donato 5, 40126 Bologna, Italy
}%
 
 \author{Mirko Degli Esposti}
\email{mirko.degliesposti@unibo.it}
\affiliation{%
Dip. Matematica, Universit\`a di Bologna, Piazza di Porta San Donato 5, 40126 Bologna, Italy
}%

\author{Georgie Knight}
 \email{georgiesamuel.knight@unibo.it}

\affiliation{%
Dip. Matematica, Universit\`a di Bologna, Piazza di Porta San Donato 5, 40126 Bologna, Italy
}%

\date{\today}

\begin{abstract}
The Sparre-Andersen theorem is a remarkable result in  one-dimensional random walk theory concerning the universality of the ubiquitous first-passage-time distribution. It states that the probability distribution $\rho_n$ of the number of steps needed for a walker starting at the origin to land on the positive semi-axes does not depend on the details of the distribution for the jumps of the walker, provided this distribution is symmetric and continuous, where in particular $\rho_n \sim n^{-3/2}$ for large number of steps $n$.  On the other hand, there are many physical situations in which the time spent by the walker in doing one step depends on the length of the step and the interest concentrates on the time needed for a return, not on the number of steps. Here we modify the Sparre-Andersen proof to deal with such cases, in rather general situations in which the time variable correlates with the step variable. As an example we present a natural process in 2D that shows deviations from normal scaling are present for the first-passage-time distribution on a semi plane. 
\end{abstract}
\maketitle

For more than a century (see, for instance \cite{Ein}) random walks have played a crucial role as a theoretical tool to model an impressive number of physical (and not only) problems. 
Fundamental questions in the theory of stochastic processes are related to the problem of when a variable in a system enters some a priori specified state for the first time: the {\em first-passage-time}. Knowledge of the first-passage-time distribution (FPTD) finds application in many diverse areas of the natural sciences and economics; from the spike distribution in neuronal dynamics, the meeting time of two molecules in diffusion-controlled chemical reactions, the cluster density in aggregation reactions, to the price of a stock reaching a specific value and the ruin problem of actuarial science (see \cite{red01} for an extensive treatment of the problem, and early references to applications); in the last decades, relevance of such a theory to non-equilibrium problems has been exploited \cite{maj-ap}. In this framework the Sparre-Andersen theorem (SA) \cite{SA54,FeBookII,KS} plays an outstanding role: in physics it has been invoked in the study of persistence in stochastic spin models and random walks \cite{Baldarassi} the study of polymer dynamics \cite{Zoia} and in the analysis of scattering from a Lorentz slab \cite{Larralde}. In particular, SA states that the probability that a random walker who starts at the origin, enters the positive semi-axis for the first time (its first-passage-time) after $n$ steps is independent of the particular details of the jump length distribution,  provided that it is symmetric about the origin and continuous: in such a case the decay of the FPTD has the universal asymptotics $n^{-3/2}$. Here the conceptual import of SA is apparent: it provides an outstanding example of {\em universality} in the realm of stochastic processes, with a huge universality class. The origin of such universality is subtler than other examples in probability: to our knowledge it cannot be encompassed by simple renormalization, like the central limit theorem (see, for example, \cite{Sorn}). 

However, in most physical situations, spatio-temporal correlations exist in the time taken to perform a jump of a certain length, a noteworthy example being represented by  L\'evy glasses \cite{Bertolotti}; an optical (quenched) L\'evy walk pinball, where the number of scattering events $n$ is not an accessible quantity, while the relevant variable is the (continuous) physical time $t$. Here we  reconsider SA in order to take into account such correlations in a continuous-time-random-walk (CTRW)  jump model \cite{variCTRW} (not relying on subordinations schemes, see \cite{KS,MK}). We hope and expect that this will foster wider applications of this beautiful result from probability theory.

The time costs associated to each jump event we consider are quite general and include the case where they correlate only to the length of a jump (as in a velocity model). Extending to our setting the method in Feller \cite{FeBookII}  (Chapter $XII$) we derive the Laplace transform of the distribution of first entry time into the positive semi-axis. To illustrate the importance of taking the real time of the process into account, we introduce a simple example of a random walk in the plane where the   universal SA scaling does not hold in continuous time, while being valid for the discrete time. In particular the FPTD does not decay like $t^{-3/2}$. We finally comment on the results for Levy Walks. 

Consider a random walker on the real line starting at the origin and  let
 $(X_1, T_1)$, $(X_2,T_2)$,...,$(X_j,T_j)$,... be a sequence of independent identically distributed (iid) \emph{pairs} of  random variables corresponding to the  steps $X$ and the associated  times $T$. More precisely, at the event $i$ the random walker waits for a time  $T_i$ after which an instantaneous jump of length  $X_i$ takes place. Denote the joint probability density of $X$ and $T$ by  $p(x,t)$. After $n$ steps the walker will be in position $S_n$ at time $C_n$ where
\begin{eqnarray}
                 &S_0=0,&\,\,S_n=\sum_{k=0}^{n}X_k \\
                 &C_0=0,&\,\,C_n=\sum_{k=0}^nT_k \, .
\label{Eq:PartialSums}
\end{eqnarray}
We can think of the random variables $C$ as a `time-cost' associated to the process. Note that, while we assume the pairs to be iid, we do not impose a priori that $T_i$ is independent of $X_i$ (i.e. in general $p(x,t)$ does not factorize), as per usual in CTRW models.

\vspace{0.3cm}
We are interested in the probability distribution of the first entrance into the positive axis, {\em i.e.} what is termed in the mathematical literature as the first {\em ladder time}.
In the (discrete-time) watch ruled by events this will be the probability $\rho_n$  that $n$ is the first index such that  $S_n>0$. In this framework, SA theorem  states that, if the distribution of jumps is continuous and symmetric, then  the generating function of $\rho_n$ is 
\begin{equation}\label{SA}
\sum_{n=1}^{\infty}{\rho_n z^n}=1-\sqrt{1-z}.
\end{equation}
Here we are interested in a continuous time setting, so the main quantity under investigation will be
$\pi(t)$, namely  the probability density for the first  ladder time. 
\begin{equation}
\label{t-n}
\pi(t)=\sum_{n=1}^\infty \, \pi_n(t),
\end{equation}
where $\pi_n(t)$ denotes the probability density to land in the positive semi-axis 
at time t for the first time after the $n$-th ``event" took place. We will provide a closed formula for the  Laplace transform of such a distribution
\begin{eqnarray}\label{zzEq:characteristicfunction}
               \hat{\pi}(s)&=&\int_{0}^{\infty} \,  \pi(t) \,e^{-st}  \,dt \\
               &=&\sum_{n=1}^\infty \, \int_0^\infty \, \pi_n(t)\, e^{-st}\,dt = \sum_{n=1}^\infty \, \hat{\pi}_n(s).\nonumber
\end{eqnarray}

Before entering into the derivation of a general expression for $\hat{\pi}(s)$ we firstly consider the simplified case $p(x,t)=r(x)\psi(t)$, where  step lengths and time costs are uncorrelated.
\begin{eqnarray}%
\label{separate}
\hat{\pi}_n(s)=\rho_n \hat{\psi}(s)^n 
\end{eqnarray}
 and thus, provided $r(x)$ is continuous and symmetric, eq Eq.(\ref{SA}) implies  
\begin{eqnarray}\label{uncorrelated}
\hat{\pi}(s)=1-\sqrt{1-\hat{\psi}(s)}.
\end{eqnarray}
This result is not surprising in such a case: when the time is independent of the jump, the effective time-cost of a step is ruled by a random (subordinated) clock  that effectively replaces the role of the simple clock in the standard SA result (see also  \cite{SokMet04}).

In the general case the factorization in Eq.(\ref{separate}) is not a priori valid and more care should be taken.
There exists a number of techniques by which the general problem can be handled, such as the Wiener-Hopf factorization scheme: it turns out that for our purposes a particularly simple procedure is via a combinatorial lemma (closely following the approach of  \cite{FeBookII}). 
Indeed the crucial  observation is that,  thanks to  its combinatorial nature, such a lemma can be equally applied to the subset of all paths that have the same time-cost, once such a time-cost is invariant under permutations of the jumps in the given path, as in our setting. A derivation of a general expression for $\hat{\pi}(s)$  thus passes through  grouping the set of exiting paths into subsets with fixed exit-time where the lemma still applies.  In the following we will show how this idea leads to resumming  the contributions of each subset and finally to an explicit expression for $\hat{\pi}(s)$.

The combinatorial lemma takes into account a generic sequence consisting of $m$ events,  a corresponding set of $m$ cyclical permutations $(X_p,..,X_m,X_1,..,X_{p-1})$ and their partial sums $S_j^{(p)}$, and states that, if $S_m>0$ then, if $r$ denotes the number of cyclically permuted sequences for which $m$ is a ladder index, then $r \geq 1$, and for each such a sequence $m$ is the $r$-th ladder index. An auxiliary set of binary random variables is then defined $\{\varepsilon_{[k];m}^{(v)}\}$, in such a way that $\varepsilon_{[k];m}^{(v)}=1$ if $m$ is the $k^{th}$ ladder index for the $v^{th}$ cyclic permutation, and $\varepsilon_{[k];m}^{(v)}=0$ otherwise. 

It is also instrumental to consider the probability densities that $t$ is the $k$-th ladder time ($k$-th absolute maximum in the sequence of partial sums): $\hat{\pi}_{[k]}(t)=\sum_m\,\hat{\pi}_{[k];m}(t)$, if, once again, we partition the physical time into ``events". Such a $k$-th ladder time distribution is useful since it appears when considering powers of the generating function (\ref{zzEq:characteristicfunction}):

\begin{equation}
\label{klad2}
\hat{\pi}(s)^j=\sum_{m=1}^{\infty}\, \int _0^\infty \, dT \, e^{-sT} \pi _{[j];m}(T),
\end{equation}
 that follows from the fact that
\begin{equation}
\pi _{[j];m}(T)= \hspace{-0.7cm}\sum_{m_1+\cdots+m_k=m} \hspace{-0.0cm}\int dt_1\cdots dt_k \delta(t-\sum_{i=1}^k t_i) \prod_ j \pi_{m_j}(t_j).\nonumber
\end{equation}
The connection with the starting issue is that
\begin{equation}
\label{pi-perm}
\pi_{[j];m}(t)=\mathrm{Prob}(\varepsilon_{[j];m}^{(1)}=1\,|\,C_m=t),
\end{equation} 
where   $\mathrm{Prob}(\cdot)$ denotes the probability density function of the corresponding variable. 
Since all the variables $\{\varepsilon^{(v)}\}$ have a common distribution, then ($\langle \cdot \rangle$ will indicate the corresponding expectation value)
\begin{equation}
\pi_{[j];m}(t)= \frac1 m \langle \varepsilon_{[j];m}^{(1)}+\cdots \varepsilon_{[j];m}^{(m)}\, | \, C_m=t \rangle,
\label{zzEq:tau-expected}
\end{equation}
and by taking into account the combinatorial lemma, we see that
\begin{equation}
\label{zzEq:tau-sum-r}
\sum_j \, \frac 1 j \pi_{[j];m}(t)=\frac 1 m \mathrm{Prob}(S_m>0\, |\,C_m=t).
\end{equation}
If we now take a sum over m, and Laplace transform (\ref{zzEq:tau-sum-r})
we obtain
\begin{equation}
         - \ln\left(1-\hat{\pi}(s)\right) =
\sum_{m=1}^{\infty} \frac{1}{m}\int_{0}^{\infty}
\mathrm{Prob}(S_m>0\, | \, C_m=t)\,e^{-st}dt.
\label{zzEq:tau-sum-generating}                
\end{equation}
Eq. (\ref{zzEq:tau-sum-generating}) is the main general result.

Under the further assumption  that $p(x,t)$ is continuous in $x$ and spatially symmetric $p(x,t)=p(-x,t)$, then 
\begin{eqnarray}
    \mathrm{Prob}(S_m>0\, | \, C_m=t)
               &=&\frac{1}{2}\mathrm{Prob}(C_m=t).\nonumber
\label{zzEq:probabilityidentity}
          \nonumber
\label{Eq:probabilityidentity}
\end{eqnarray}
As the variables $T_j$  are identically distributed (i.e. they do not depend on $j$), by the convolution theorem Eq.(\ref{zzEq:tau-sum-generating})
reduces to
\begin{equation}\label{backGF}
               \hat{\pi}(s)=1-\sqrt{1-c(s)},
\end{equation}
where 
\begin{equation}
               c(s)=\int_{-\infty}^{\infty}   \int_{0}^{\infty}  p(x,t) \,e^{-st}  \,dt dx
\label{Eq:characteristicfunction-2}
\end{equation}
is the cost-associated characteristic function. 
Note that if $p(x,t)$ factorizes then $c(s)=\hat{\psi}(s)$ thus recovering  Eq.(\ref{uncorrelated}).
\\

A particularly interesting case is when the time length is directly correlated to the step length so that we have $p(x,t)=r(x)\delta(t-|x|)$  (as in the Levy Walk,  note also the discussion at the end of the paper). Indeed it is straightforward  to observe that
\begin{itemize}
\item if steps have a finite average length $\langle l \rangle $ then for small $s$ we have $c(s)\sim 1-\langle l\rangle s$ and thus
$\pi(s)\sim 1- (\langle l \rangle s)^{1/2}$ giving the  conventional SA asymptotics $\pi(t) \sim t^{-3/2}$ via Tauberian theorems;
\item if steps have an infinite average length and their distribution asymptotically  decays as $r(x)\sim |x|^{-(1+\alpha)}$ $\alpha \in(0,1)$; then for small $s$ we have $c(s)\sim 1-Bs^{\alpha}$ and thus
$\tau(s)\sim 1- (Bs^{\alpha})^{1/2}$ giving the asymptotic $t^{-(1+\alpha/2)}$, similarly to the subdiffusive uncorrelated case (see for example \cite{Korabel, Dybiec}, and \cite{B,MeK}).
\end{itemize}

Now, we will use Eq.(\ref{backGF}) to discuss a process   in 2D that shows deviations from the SA scaling for the first-passage-time-distribution on a semi plane. 
At each step, a vertical barrier is chosen at a random $x$ position drawn from a distribution $r(x)$ on $\rz$. When hitting the barrier  the particle is scattered with  uniform outgoing angle $\psi$, while maintaining unit velocity: then a new random barrier is placed.  We will derive the FPTD on the negative semi-plane for a particle starting at the origin. Note that such a problem is solved by just projecting on the $x$-axis. The physical time associated with each step is  $l/\cos(\psi)$, we can use  Eq.(\ref{backGF}) to study the distribution of first entry times  into the region $x>0$ in this setting. 
From a purely one-dimensional perspective this is a model with a fluctuating velocity: other types of fluctuating velocity have been discussed in \cite{Denisov}.
The outgoing angle introduces a further random variable $\xi=\frac1{|v_x|}=\frac1{\cos \psi}$ so our starting object is (assuming unit velocity)
\begin{equation}
\label{p-xi}
p_\xi(x,t)=\phi(|x|)\delta(t-|x|\xi)
\end{equation}
In order to compute the asymptotic behavior of the FPTD we have to evaluate
\begin{equation}
\label{c-xi}
c(s)=\int_0^\infty \, dl \, \phi(l) \, \int_1^\infty\, d \xi  \,\wp(\xi)\,\int_0^\infty \, dt \, e^{-st}\,\delta(t-l\xi).
\end{equation}
Since our interest is in investigating the anomalies induced by the $\xi$ distribution, we take the simplest possible step-length distribution $\phi_{\spadesuit}(l)=\delta(l-1)$. The generating function thus simplifies to
\begin{equation}
\label{c-xi-simp}
c_{\spadesuit}(s)=\frac 2{\pi}\int_1^\infty\, d\xi \,e^{-s \xi} \, \frac 1{\xi \sqrt{\xi^2-1}},
\end{equation}
and we are interested in the small $s$ asymptotics: this is easily obtained if we notice that
\begin{equation}
\label{dc-xi-simp}
\frac{dc_{\spadesuit}(s)}{ds}=-\frac 2 \pi \int_1^\infty\, d\xi \,e^{-s \xi}\,\frac 1 {\sqrt{\xi^2-1}}=-\frac 2 \pi K_0(s), 
\end{equation}
where $K_0(s)$ is a modified Bessel function. Since  for small~$ s$
\begin{equation}
\label{smallK}
K_0(s) \sim -\ln s
\end{equation}
we have that,
\begin{equation}
\label{c-xi-fin}
c_{\spadesuit}(s) \sim 1+s \ln s
\end{equation}
thus implying a logarithmic correction  $\pi(t) \sim  (\ln t)^{\frac12} t^{-\frac32} $ for large $t$ again via Tauberian theorems. It is easy to check that the logarithmic correction appears for {\emph{any}} choice of step length distribution $\phi(l)$.
Finally, we may verify that our calculations remain essentially unaltered if we  go to a 3D setting, where parallel scattering planes are distributed according to $\phi(d)$, and each scattering event result in a uniformly distributed outgoing angle. Take for instance planes parallel to $\pi_{xy}$, so that $\xi=1/\cos \theta$. Then we have
\begin{equation}
\label{distan3}
\wp(\theta)=\frac 12\sin \theta \qquad \theta \in [0, \pi/2]
\end{equation}
and, by following former steps, we get, through a change of variable
\begin{equation}
\label{dis-xi3}
\wp(\xi)=\frac 1 {\xi^2} \qquad \xi \in [1, \infty),
\end{equation}
leading to
\begin{equation}
\label{c3}
c_{\spadesuit}^{(3)}(s)=\int_1^\infty\, d\xi \, e^{-s\xi} \, \frac 1 {\xi^2}=E_2(s)
\end{equation}
where $E_2$ is an exponential integral function, and
again the small $s$ expression is 
\begin{equation}
\label{c3s}
c_{\spadesuit}^{(3)}(s) \sim 1 +s \ln s.
\end{equation}

To conclude, we compute  the FPTD for a walker not starting at the origin but at a random position  $X_0$,  distributed as all other $X$. This is equivalent to saying that at the $i$-th event the random walker first jumps a distance given by $X_i$ and then waits for a time given by $T_i$, differently from the models introduced before, where the jump is considered completed  only \emph{after} the associated time-cost is passed. If we let the  $T_j$ be distributed independently of the step-lengths $X_j$ we have, similarly to Eq.( \ref{separate}),  that $\hat{\pi}_n(s)=\rho_n \hat{\psi}(s)^{n-1}$ and thus 
 $              \hat{\pi}(s)=\frac{1}{\hat{\psi}(s)}\left[1-\sqrt{1-\hat{\psi}(s)}\right]$
that is somehow surprising given that the Pollatzeck-Spitzer formula is strongly dependent on the starting point $x_0$ \cite{Maj2010}.
As an illustrative example, consider a discrete-time one-dimensional random walker with jumps taken uniformly on $[-1,1]$ \cite{nott} and with the time associated to each step  chosen at random, independent to the step, to be equal to $1$ or $2$ with equal probability. That is we have $p(x,n)=r(x)q(n)$ where
\begin{eqnarray}
r(x)&=& \frac{1}{2} \quad \textrm{for} \, x\in [-1,1]\nonumber\\
q(n)&=&\frac{1}{2}\left[\delta(n-1)+\delta(n-2)\right].
\end{eqnarray}
If we start the random walker from  $X_0=0$, corresponding to the wait-then-jump model we have from  Eq.(\ref{backGF}) the generating function  (that takes the role of the Laplace transform in discrete time) is then given by
\begin{equation}\label{gf-example1}
               \hat{\pi}(z)=\left[1-\sqrt{1-\frac1{2} [ z+ z^2]}\right],
\end{equation} 
whereas if we start the random walker at a random initial position distributed as $X$, corresponding to the jump-then-wait model we have 
\begin{equation}\label{gf-example2}
               \hat{\pi}(z)=\frac{2}{(z+z^2)}\left[1-\sqrt{1-\frac1{2} [ z+ z^2]}\right].
\end{equation}

Note that the  two models presented can be used to study  the first passage time problem  for the CTRW velocity model, where the random walker moves with a constant velocity between steps. More precisely, for this model we investigate the first time the walker traverses the origin rather than the first time a scattering event occurs  in the positive semi-axes. Indeed, given the distribution of jumps and velocities  defining the CTRW velocity model, there are two obvious related    jump-then-wait and wait-then-jump  models associate with it, where the corresponding walkers respectively always precedes and follows the  velocity-model walker. Let $t_w$ be the time of first passage for the wait-then-jump walker, let $t_j$ be the time of the first passage of the jump-then-wait walker and let $t_c$ be the time of first crossing of the continuous time walker. For every given walk one has $t_j\leq t_c \leq t_w$, which implies $\mathbf{P}\{ t_j>t\} \leq \mathbf{P}\{ t_c>t\} \leq \mathbf{P}\{ t_w>t\}$. 

For a CTRW velocity model with independent step lengths and velocities, (provided  the corresponding average time-cost is finite) the time-asymptotic behavior for the two bounding processes is the same and of SA type $(\sim t^{-\frac32})$  and thus we can conclude that  the decay of the first return time for the velocity model is  also of  SA type in this case. Whether SA scaling is found for the CTRW velocity model with spatio-temporal correlations (as for example for a bona fide Levy Walk) or for more general systems such as those with correlations between the step-times  (as in \cite{Chetal09}) remains an open question.
\begin{acknowledgments}
We acknowledge partial support by the FIRB-project RBFR08UH60 (MIUR, Italy).
\end{acknowledgments}


\begin{thebibliography}{17}%
\makeatletter
\providecommand \@ifxundefined [1]{%
 \@ifx{#1\undefined}
}%
\providecommand \@ifnum [1]{%
 \ifnum #1\expandafter \@firstoftwo
 \else \expandafter \@secondoftwo
 \fi
}%
\providecommand \@ifx [1]{%
 \ifx #1\expandafter \@firstoftwo
 \else \expandafter \@secondoftwo
 \fi
}%
\providecommand \natexlab [1]{#1}%
\providecommand \enquote  [1]{``#1''}%
\providecommand \bibnamefont  [1]{#1}%
\providecommand \bibfnamefont [1]{#1}%
\providecommand \citenamefont [1]{#1}%
\providecommand \href@noop [0]{\@secondoftwo}%
\providecommand \href [0]{\begingroup \@sanitize@url \@href}%
\providecommand \@href[1]{\@@startlink{#1}\@@href}%
\providecommand \@@href[1]{\endgroup#1\@@endlink}%
\providecommand \@sanitize@url [0]{\catcode `\\12\catcode `\$12\catcode
  `\&12\catcode `\#12\catcode `\^12\catcode `\_12\catcode `\%12\relax}%
\providecommand \@@startlink[1]{}%
\providecommand \@@endlink[0]{}%
\providecommand \url  [0]{\begingroup\@sanitize@url \@url }%
\providecommand \@url [1]{\endgroup\@href {#1}{\urlprefix }}%
\providecommand \urlprefix  [0]{URL }%
\providecommand \Eprint [0]{\href }%
\providecommand \doibase [0]{http://dx.doi.org/}%
\providecommand \selectlanguage [0]{\@gobble}%
\providecommand \bibinfo  [0]{\@secondoftwo}%
\providecommand \bibfield  [0]{\@secondoftwo}%
\providecommand \translation [1]{[#1]}%
\providecommand \BibitemOpen [0]{}%
\providecommand \bibitemStop [0]{}%
\providecommand \bibitemNoStop [0]{.\EOS\space}%
\providecommand \EOS [0]{\spacefactor3000\relax}%
\providecommand \BibitemShut  [1]{\csname bibitem#1\endcsname}%
\let\auto@bib@innerbib\@empty
\bibitem{Ein} S. Chandrasekhar, Rev. Mod. Phys. {\bf 15}, 1 (1943).
\bibitem [{\citenamefont {Redner}(2001)}]{red01}%
  \BibitemOpen
  \bibfield  {author} {\bibinfo {author} {\bibfnamefont {S.}~\bibnamefont
  {Redner}},\ }\href {http://books.google.it/books?id=xtsqMh3VC98C} {\emph
  {\bibinfo {title} {A Guide to First-Passage Processes}}}\ (\bibinfo
  {publisher} {Cambridge University Press},\ \bibinfo {year}
  {2001})\BibitemShut {NoStop}%
%
%
%
%
%
%
\bibitem{maj-ap} A.J. Bray, S.N. Majumdar and G. Schehr, Adv. Phys. {\bf 62}, 225 (2013).
\bibitem [{\citenamefont {Andersen}(1954)}]{SA54}%
  \BibitemOpen
  \bibfield  {author} {\bibinfo {author} {\bibfnamefont {E.}\ \bibnamefont
  {Sparre Andersen}},\ }\href {http://eudml.org/doc/165543} {\bibfield  {journal}
  {\bibinfo  {journal} {Math. Scand.}\ }\textbf {\bibinfo {volume} {2}},\
  \bibinfo {pages} {195} (\bibinfo {year} {1954})}\BibitemShut {NoStop}%
%
\bibitem [{\citenamefont {Feller}(1971)}]{FeBookII} \BibitemOpen  \bibfield  {author} {\bibinfo {author} {\bibfnamefont {W.}~\bibnamefont  {Feller}},\ }\href {http://books.google.it/books?id=BsSwAAAAIAAJ} {\emph   {\bibinfo {title} {An Introduction to Probability Theory and Its Applications   II}}}\ (\bibinfo  {publisher} {Wiley},\ \bibinfo {year} {1971})\BibitemShut   {NoStop}%
\bibitem{KS} J. Klafter and I.~M. Sokolov, {\it First steps in random walks} (Oxford University Press, 2011).
%
%
%
\bibitem [{\citenamefont {Baldarassi}\ \emph {et~al.}(1999)\citenamefont
  {Baldarassi}}]{Baldarassi}%
  \BibitemOpen
  \bibfield  {author} {\bibinfo {author} {\bibfnamefont {A.}~\bibnamefont
  {Baldarassi}}, \bibinfo {author} {\bibfnamefont {J.~P.}~\bibnamefont {Bouchard}},
  \bibinfo {author} {\bibfnamefont {I.}~\bibnamefont {Dornic}} \ and\
  \bibinfo {author} {\bibfnamefont {C.}~\bibnamefont {Godr\'{e}che}},\ }\href
  {\doibase10.1103/PhysRevE.59.R20} {\bibfield  {journal} {\bibinfo
  {journal} {Phys. Rev. E}\ }\textbf {\bibinfo {volume} {59}},\ \bibinfo
  {pages} {R20--R23} (\bibinfo {year} {1999})}; \bibfield  {author} {\bibinfo {author} {\bibfnamefont {M.}~\bibnamefont
  {Bauer}}, \bibinfo {author} {\bibfnamefont {C.}~\bibnamefont {Godr\'{e}che}}\ and\
  \bibinfo {author} {\bibfnamefont {J.~M.}~\bibnamefont {Luck}},\ }\href
  {\doibase10.1023/A:1004636216365} {\bibfield  {journal} {\bibinfo
  {journal} {J. Stat. Phys.}\ }\textbf {\bibinfo {volume} {96}},\ \bibinfo
  {pages} {963-1019} (\bibinfo {year} {1999})}\BibitemShut {NoStop}%
%
%
\bibitem [{\citenamefont {Zoia}\ \emph {et~al.}(2009)\citenamefont
  {Zoia}, \citenamefont {Rossman},\ and\ \citenamefont
  {Majumdar}}]{Zoia}%
  \BibitemOpen
  \bibfield  {author} {\bibinfo {author} {\bibfnamefont {A.}\ \bibnamefont
  {Zoia}}, \bibinfo {author} {\bibfnamefont {A.}~\bibnamefont {Rosso}}, \
  and\ \bibinfo {author} {\bibfnamefont {S.~N.}\ \bibnamefont {Majumdar}},\
  }\href {\doibase 10.1103/PhysRevLett.102.120602} {\bibfield  {journal} {\bibinfo
  {journal} {Phys. Rev. Lett.}\ }\textbf {\bibinfo {volume} {102}},\ \bibinfo
  {pages} {120602} (\bibinfo {year} {2009})}\BibitemShut {NoStop}%
%
%
\bibitem [{\citenamefont {Larralde}\ \emph {et~al.}(1998)\citenamefont
  {Larralde}, \citenamefont {Leyvraz}, \citenamefont {Martinez-Mekler},
  \citenamefont {Rechtman},\ and\ \citenamefont {Ruffo}}]{Larralde}%
  \BibitemOpen
  \bibfield  {author} {\bibinfo {author} {\bibfnamefont {H.}~\bibnamefont
  {Larralde}}, \bibinfo {author} {\bibfnamefont {F.}~\bibnamefont {Leyvraz}},
  \bibinfo {author} {\bibfnamefont {G.}~\bibnamefont {Martinez-Mekler}},
  \bibinfo {author} {\bibfnamefont {R.}~\bibnamefont {Rechtman}}, \ and\
  \bibinfo {author} {\bibfnamefont {S.}~\bibnamefont {Ruffo}},\ }\href
  {\doibase 10.1103/PhysRevE.58.4254} {\bibfield  {journal} {\bibinfo
  {journal} {Phys. Rev. E}\ }\textbf {\bibinfo {volume} {58}},\ \bibinfo
  {pages} {4254} (\bibinfo {year} {1998})}\BibitemShut {NoStop}%
%
\bibitem{Sorn} D. Sornette, {\it Critical phenomena in natural sciences, 2${}^{nd}$ edition} (Springer, 2006).
%
\bibitem [{\citenamefont {Pierre~Barthelemy}\ and\ \citenamefont
  {Wiersma}(2008)}]{Bertolotti}%
\BibitemOpen
  \bibfield  {author} {\bibinfo {author} {\bibfnamefont {P.~Barthelemy},\ \bibnamefont
  {J.~Bertolotti}}\ and\ \bibinfo {author} {\bibfnamefont {D.~S.}\
  \bibnamefont {Wiersma}},\ }\href {http://dx.doi.org/10.1038/nature06948}
  {\bibfield  {journal} {\bibinfo  {journal} {Nature}\ }\textbf {\bibinfo
  {volume} {453}},\ \bibinfo {pages} {495} (\bibinfo {year}
  {2008})}\BibitemShut {NoStop}%
%
%
\bibitem [{\citenamefont {Montroll}\ and\ \citenamefont
  {Weiss}(1965)}, {\citenamefont {Klafter},\\citenamefont
  {Klafter}, \citenamefont {Blumen},\ and\ \citenamefont
  {Shlesinger}(1987)}, {\citenamefont {Klafter},\\citenamefont
  {Klafter}, \citenamefont {Blumen},\ and\ \citenamefont
  {Shlesinger}(1987)}]{variCTRW}%
  \BibitemOpen
  \bibfield  {author} {\bibinfo {author} {\bibfnamefont {E.~W.}\ \bibnamefont
  {Montroll}}\ and\ \bibinfo {author} {\bibfnamefont {G.~H.}\ \bibnamefont
  {Weiss}},\ }\href {\doibase 10.1063/1.1704269} {\bibfield  {journal}
  {\bibinfo  {journal} {J. Math. Phys.}\ }\textbf {\bibinfo {volume} {6}},\
  \bibinfo {pages} {167} (\bibinfo {year} {1965})}\BibitemShut {NoStop}; %
   \bibfield  {author} {\bibinfo {author} {\bibfnamefont {J.}~\bibnamefont
  {Klafter}}, \bibinfo {author} {\bibfnamefont {A.}~\bibnamefont {Blumen}},
  \ and\ \bibinfo {author} {\bibfnamefont {M.~F.}~\bibnamefont {Shlesinger}},\
  }\href {\doibase 10.1103/PhysRevA.35.3081} {\bibfield  {journal} {\bibinfo
  {journal} {Phys. Rev. A}\ }\textbf {\bibinfo {volume} {35}},\ \bibinfo
  {pages} {3081} (\bibinfo {year} {1987})}\BibitemShut {NoStop}; %
  \bibfield  {author} {\bibinfo {author} {\bibfnamefont {G.}\ \bibnamefont
  {Zumofen}}\ and\ \bibinfo {author} {\bibfnamefont {J.}~\bibnamefont
  {Klafter}},\ }\href {\doibase 10.1103/PhysRevE.47.851}
  {\bibfield  {journal} {\bibinfo  {journal} {Phys. Rev. E}\
  }\textbf {\bibinfo {volume} {47}},\ \bibinfo {pages} {851} (\bibinfo {year}
  {1993})}\BibitemShut {NoStop}%
%
%
\bibitem{MK} R. Metzler and J. Klafter, Phys. Rep. {\bf 339},1 (2000).
\bibitem [{\citenamefont {Sokolov}\ and\ \citenamefont
  {Metzler}(2004)}]{SokMet04}%
  \BibitemOpen
  \bibfield  {author} {\bibinfo {author} {\bibfnamefont {I.~M.}\ \bibnamefont
  {Sokolov}}\ and\ \bibinfo {author} {\bibfnamefont {R.}~\bibnamefont
  {Metzler}},\ }\href {http://stacks.iop.org/0305-4470/37/i=46/a=L02}
  {\bibfield  {journal} {\bibinfo  {journal} {J. Phys. A: Math. Gen.}\
  }\textbf {\bibinfo {volume} {37}},\ \bibinfo {pages} {L609} (\bibinfo {year}
  {2004})}\BibitemShut {NoStop}%
%
%
\bibitem [{\citenamefont {Korabel}\ and\ \citenamefont
  {Barkai}(2011)}]{Korabel}%
  \BibitemOpen
  \bibfield  {author} {\bibinfo {author} {\bibfnamefont {N.}~\bibnamefont
  {Korabel}}\ and\ \bibinfo {author} {\bibfnamefont {E.}~\bibnamefont
  {Barkai}},\ }\href {http://stacks.iop.org/1742-5468/2011/i=05/a=P05022}
  {\bibfield  {journal} {\bibinfo  {journal} {J. Stat. Mech.: Theory E}\
  }\textbf {\bibinfo {volume} {2011}},\ \bibinfo {pages} {P05022} (\bibinfo
  {year} {2011})}\BibitemShut {NoStop}%
%
%
\bibitem [{\citenamefont {Dybiec}\ and\ \citenamefont
  {Gudowska-Nowak}(2009)}]{Dybiec}%
  \BibitemOpen
  \bibfield  {author} {\bibinfo {author} {\bibfnamefont {B.}~\bibnamefont
  {Dybiec}}\ and\ \bibinfo {author} {\bibfnamefont {E.}~\bibnamefont
  {Gudowska-Nowak}},\ }\href {\doibase 10.1209/0295-5075/88/10003} {\bibfield
  {journal} {\bibinfo  {journal} {EPL}\ }\textbf {\bibinfo {volume} {88}},\
  \bibinfo {pages} {10003} (\bibinfo {year} {2009})}\BibitemShut {NoStop}%
 %
 \bibitem{B} E. Barkai, Phys. Rev. E {\bf 63}, 046118 (2001).
 \bibitem{MeK} R. Metzler and J. Klafter, Physica A {\bf 278}, 107 (2000).
%
\bibitem [{\citenamefont {Denisov}\ \emph {et~al.}(2012)\citenamefont
  {Denisov}, \citenamefont {Zaburdaev},\ and\ \citenamefont
  {H\"anggi}}]{Denisov}%
  \BibitemOpen
  \bibfield  {author} {\bibinfo {author} {\bibfnamefont {S.}~\bibnamefont
  {Denisov}}, \bibinfo {author} {\bibfnamefont {V.}~\bibnamefont {Zaburdaev}},
  \ and\ \bibinfo {author} {\bibfnamefont {P.}~\bibnamefont {H\"anggi}},\
  }\href {\doibase 10.1103/PhysRevE.85.031148} {\bibfield  {journal} {\bibinfo
  {journal} {Phys. Rev. E}\ }\textbf {\bibinfo {volume} {85}},\ \bibinfo
  {pages} {031148} (\bibinfo {year} {2012})}\BibitemShut {NoStop}%
%
%
\bibitem [{\citenamefont {Majumdar}(2010)}]{Maj2010}%
  \BibitemOpen
  \bibfield  {author} {\bibinfo {author} {\bibfnamefont {S.~N.}\ \bibnamefont
  {Majumdar}},\ }\href {\doibase 10.1016/j.physa.2010.01.021} {\bibfield
  {journal} {\bibinfo  {journal} {Physica A}\ }\textbf {\bibinfo {volume}
  {389}},\ \bibinfo {pages} {4299 } (\bibinfo {year} {2010})}.
  \bibfield  {author} {\bibinfo {author} {\bibfnamefont {F.}~\bibnamefont
  {Pollaczek}},\ }\href@noop {} {\bibfield  {journal} {\bibinfo  {journal}
  {Comptes Rendus}\ }\textbf {\bibinfo {volume} {234}},\ \bibinfo {pages}
  {2334} (\bibinfo {year} {1952})};
  \bibfield  {author} {\bibinfo {author} {\bibfnamefont {F.}~\bibnamefont
  {Spitzer}},\ }\href@noop {} {\bibfield  {journal} {\bibinfo  {journal}
  {Trans. Amer. Math. Soc}\ }\textbf {\bibinfo {volume} {82}},\ \bibinfo
  {pages} {323} (\bibinfo {year} {1956})};
  \bibfield  {author} {\bibinfo {author} {\bibfnamefont {F.}~\bibnamefont
  {Spitzer}},\ }\href@noop {} {\bibfield  {journal} {\bibinfo  {journal} {Duke
  Math. J.}\ }\textbf {\bibinfo {volume} {24}},\ \bibinfo {pages} {327}
  (\bibinfo {year} {1957})}\BibitemShut {NoStop}%
%
%
\bibitem{nott} Note that the choice of the distribution for the jumps is irrelevant as long as it is symmetric and continuous around $x=0$.
\bibitem [{\citenamefont {Chechkin}\ \emph {et~al.}(2009)\citenamefont
  {Chechkin}, \citenamefont {Hofmann},\ and\ \citenamefont
  {Sokolov}}]{Chetal09}%
  \BibitemOpen
  \bibfield  {author} {\bibinfo {author} {\bibfnamefont {A.~V.}\ \bibnamefont
  {Chechkin}}, \bibinfo {author} {\bibfnamefont {M.}~\bibnamefont {Hofmann}}, \
  and\ \bibinfo {author} {\bibfnamefont {I.~M.}\ \bibnamefont {Sokolov}},\
  }\href {\doibase 10.1103/PhysRevE.80.031112} {\bibfield  {journal} {\bibinfo
  {journal} {Phys. Rev. E}\ }\textbf {\bibinfo {volume} {80}},\ \bibinfo
  {pages} {031112} (\bibinfo {year} {2009})}\BibitemShut {NoStop}%
\end{thebibliography}

%

\end{document}